\documentclass{ws-ijmpa}

\begin{document}

\markboth{Roman Pasechnik, Antoni Szczurek, Oleg Teryaev} {Spin
effects in diffractive charmonia production}

%
\catchline{}{}{}{}{}
%

\title{SPIN EFFECTS IN DIFFRACTIVE CHARMONIA PRODUCTION}

\author{ROMAN PASECHNIK}

\address{Department of Physics and Astronomy,
Uppsala University,\\ Box 516, SE-751 20 Uppsala, Sweden\\
roman.pasechnik@fysast.uu.se}

\author{ANTONI SZCZUREK}

\address{Institute of Nuclear Physics PAN, PL-31-342 Cracow,
Poland and\\
University of Rzesz\'ow, PL-35-959 Rzesz\'ow, Poland\\
antoni.szczurek@ifj.edu.pl}

\author{OLEG TERYAEV}

\address{Bogoliubov Laboratory of Theoretical Physics, JINR,
Dubna 141980, Russia\\
teryaev@theor.jinr.ru}

\maketitle

\begin{history}
\received{Day Month Year}
\revised{Day Month Year}
\end{history}

\begin{abstract}
We consider exclusive double diffractive production of polarised
axial-vector $\chi_c(1^+)$ and tensor $\chi_c(2^+)$ charmonia in
proton-(anti)proton collisions at Tevatron energy. The corresponding
amplitudes for these processes are derived within the
$k_t$-factorisation approach. Contributions from different
polarisation states of axial-vector and tensor charmonia are
quantified. Corresponding experimental consequences are discussed.

\end{abstract}

\ccode{PACS numbers: 11.25.Hf, 123.1K}


\section{Introduction}

Recently, the central exclusive charmonia production has attracted a
lot of attention from both experimental [\refcite{exp}] and
theoretical [\refcite{theor}] sides. Such a process provides the
unique opportunity to test the QCD diffractive Khoze-Martin-Ryskin
(KMR) mechanism [\refcite{KMR}] based on $k_t$-factorisation
incorporating nonperturbative small-$x/k_t$ gluon dynamics described
by the unintegrated gluon distribution functions (UGDFs) against
accessible data.

The goal of the present paper is to analyze polarisation effects in
the central exclusive charmonia production. Such effects can be
identified potentially by measuring the angular distribution of
$J/\psi$ mesons from radiative decays of $\chi_c(J^+)$ giving more
detailed information on partial spin contributions. Moreover,
certain combinations of polarisation observables can be less
sensitive to unknown nonperturbative effects leading to unique
opportunities for model-independent analysis of diffractive
processes.

\section{Central exclusive production of polarised axial-vector and tensor
charmonia}

According to the Khoze-Martin-Ryskin approach (KMR) [\refcite{KMR}], we
write the amplitude of the exclusive double diffractive color
singlet production $pp\to pp\chi_{cJ}$ as
\begin{eqnarray}
{\cal
M}^{g^*g^*}_{J,\lambda}=\frac{s}{2}\frac{\pi^2\delta_{c_1c_2}}{N_c^2-1}\,\Im\int
d^2
q_{0,t}V^{c_1c_2}_{J,\lambda}\frac{f^{\text{off}}_{g,1}(x_1,x_1',q_{0,t}^2,
q_{1,t}^2,t_1)f^{\text{off}}_{g,2}(x_2,x_2',q_{0,t}^2,q_{2,t}^2,t_2)}
{q_{0,t}^2\,q_{1,t}^2\, q_{2,t}^2} \; . \label{ampl}
\end{eqnarray}
where $f^{\text{off}}_{g,i}(x_i,x_i',q_{0,t}^2,q_{i,t}^2,t_i)$ are
the off-diagonal unintegrated gluon distributions for ``active''
gluons with momenta $q_i=x_ip_i+q_{i,t}$ and color indices $c_i$,
and the screening soft gluon with small fraction $x'\ll
x_i,\,q_0\simeq q_{0,t}$, $J$ and $\lambda$ are the spin and
helicity of a produced meson with momentum $P=q_1+q_2$ and mass $M$
in the center-of-mass frame of colliding protons and $z$ axis
directed along the meson momentum ${\bf P}$, respectively.
Hard subprocess parts
$V^{c_1c_2}_{J,\lambda}(g^*g^*\to\chi_c)$ were
calculated previously in Refs.~[\refcite{chic}], and we will not shown
them explicitly here.

\begin{figure}[!h]
\begin{minipage}{0.328\textwidth}
 \centerline{\includegraphics[width=1.15\textwidth]{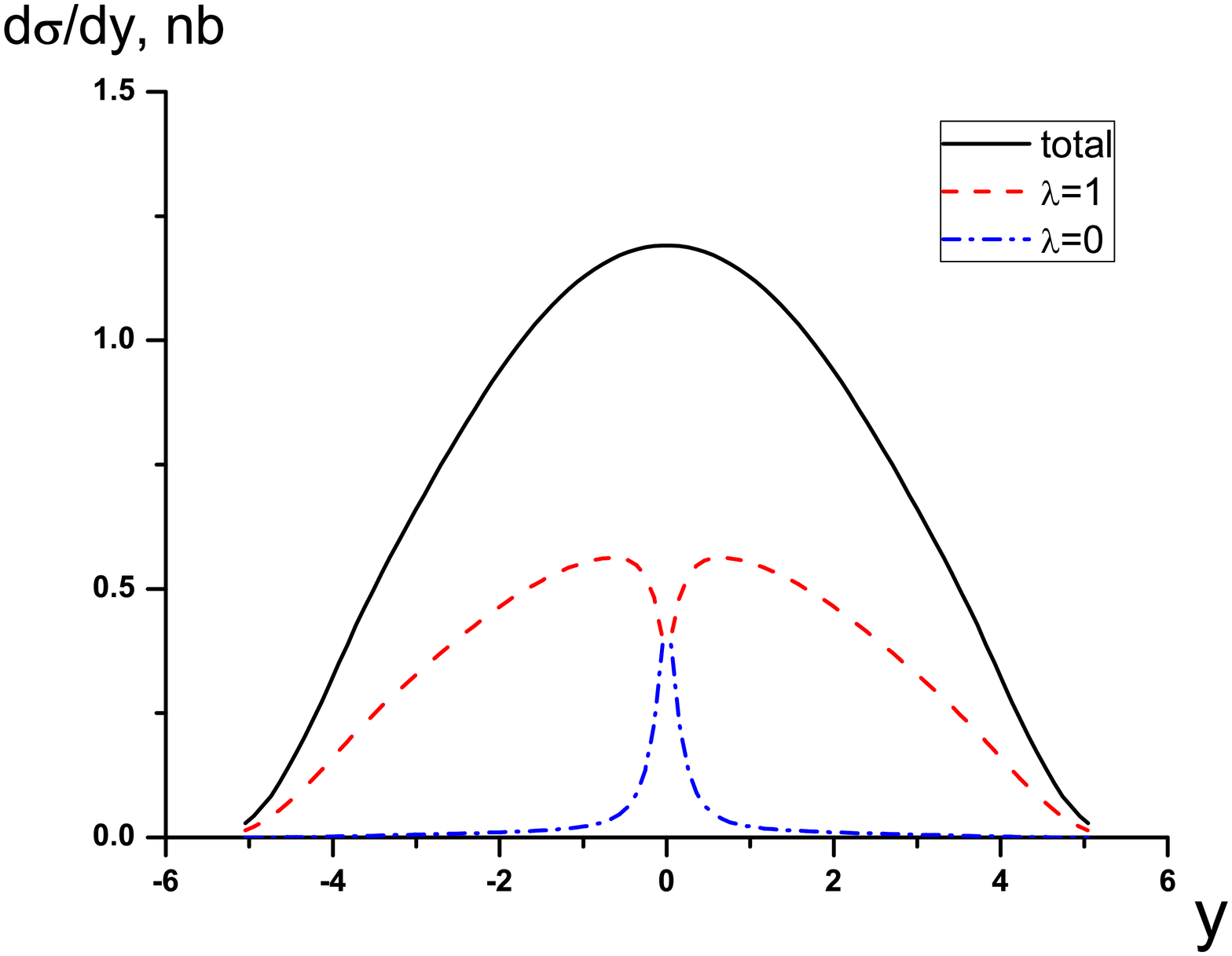}}
\end{minipage}
\begin{minipage}{0.328\textwidth}
 \centerline{\includegraphics[width=1.15\textwidth]{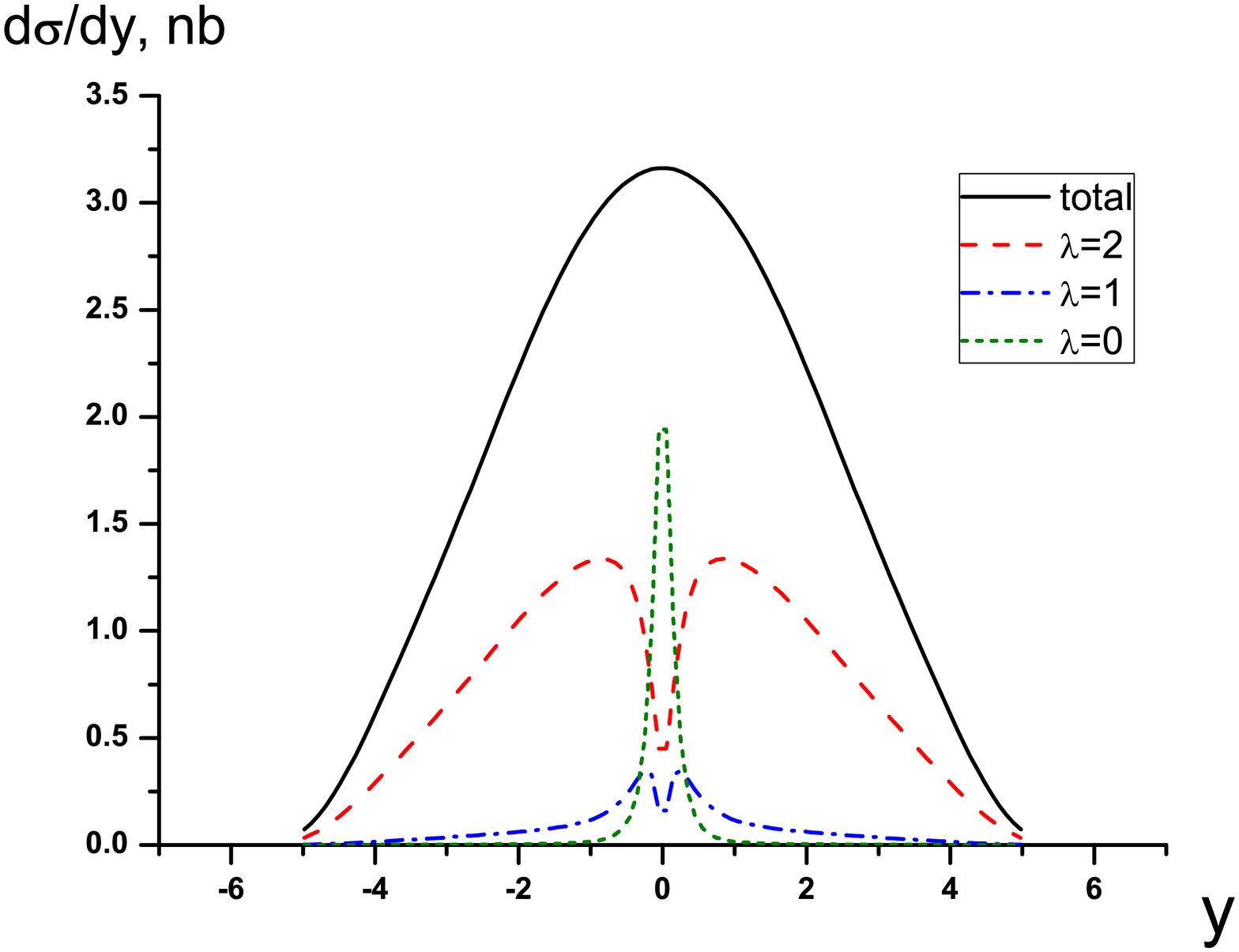}}
\end{minipage}
\begin{minipage}{0.328\textwidth}
 \centerline{\includegraphics[width=1.15\textwidth]{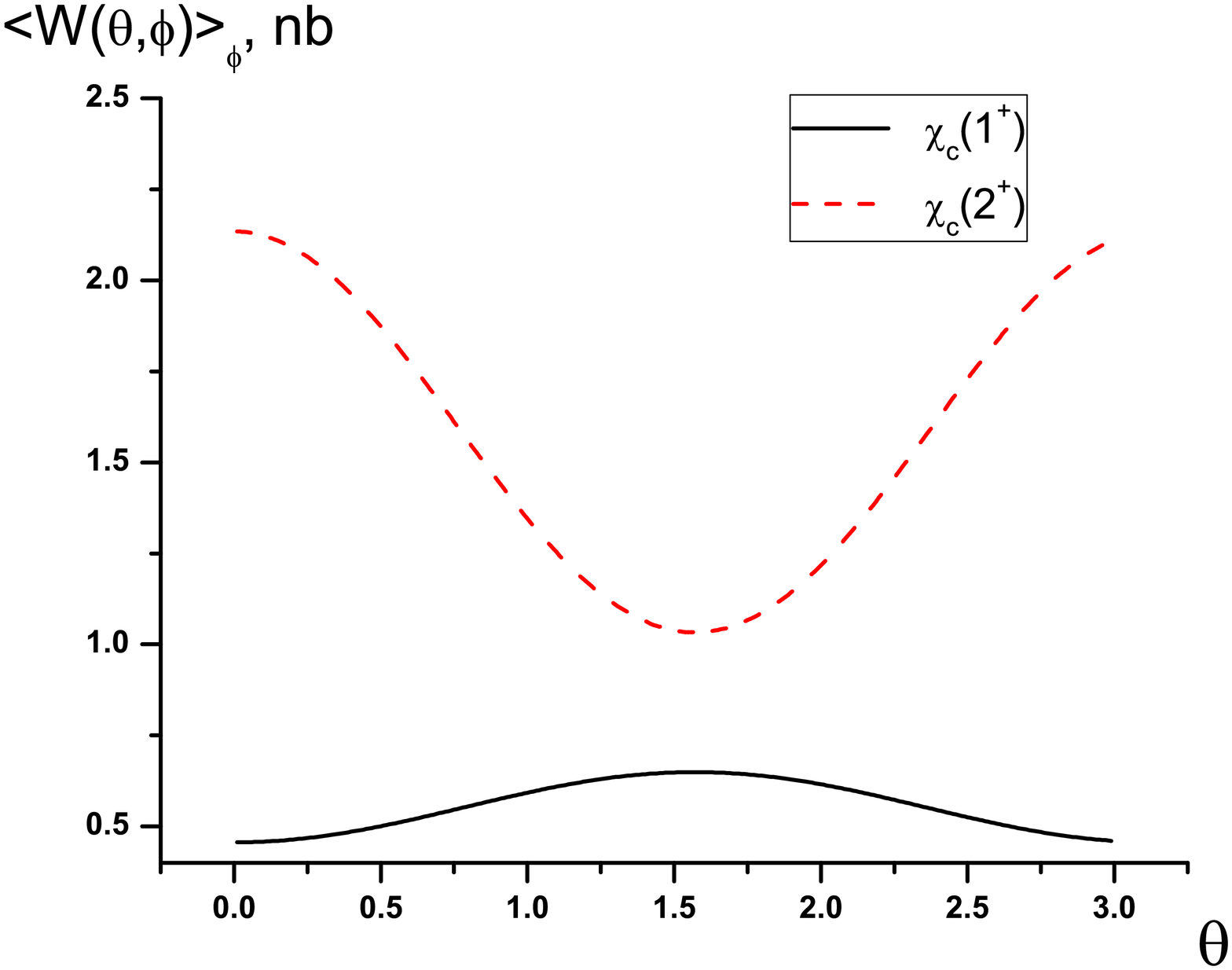}}
\end{minipage}
   \caption{\label{fig:chic}\small \em Distributions of the CEP cross section
in rapidity $y$ of $\chi_c(1^+)$ (left) and $\chi_c(2^+)$ (middle)
mesons for different polarisation states. In the right panel the
distribution of outgoing $J/\psi$ meson in the polar angle $\theta$
averaged over azimuthal angle $\phi$ is also given for $\chi_c(1^+)$
(solid line) and $\chi_c(2^+)$ (dashed line) mesons. KMR UGDF is
used.}
\end{figure}

In Fig.~\ref{fig:chic} we show differential distributions of the
central exclusive $\chi_c(1^+)$ (left) and $\chi_c(2^+)$ (right)
production cross section in rapidity $y$ for different meson
polarisations $\lambda =0,\,\pm1,\,\pm2$ (calculated with KMR UGDF
[\refcite{MR}]). They exhibit maxima/minima in the central rapidity
region $y\sim 0$. Interestingly enough, these maxima/minima in
partial helicity contributions cancel each other in the total
(summed over all meson helicity states) cross section, which has a
regular and smooth behavior around $y\to 0$.
The cross section integrated over all possible meson rapidities $y$
(in our case $|y|\leq 6.0$) is dominated by maximal meson helicity
contributions, i.e. by $|\lambda|=1$ for $\chi_c(1^+)$ and by
$|\lambda|=2$ for $\chi_c(2^+)$. This confirms that the appearance
of non-maximal helicities is a kinematical effect absent in the
spin-averaged cross-section. We do not take into account the
absorbtive effects here as they can, in principle, differ for
different meson polarisations.

\section{Angular distribution of $J/\psi$ meson}

Let us consider the central exclusive production process $pp \to
pp\chi_{cJ}$ followed by the radiative decays $\chi_{cJ}\to
J/\psi+\gamma$. Below we follow notations in Ref.~[\refcite{Kniehl03}].
Let $\theta$ and $\phi$ be the polar and azimuthal angles of the
$J/\psi$ meson in the respective coordinate system of $\chi_c(J^+)$
rest frame (this is so-called helicity frame). Then, the angular
distribution of the $J/\psi$ meson from $\chi_c(1^+)$ meson in the
general form reads
\begin{align}\nonumber
&W^{J=1}(\theta,\phi)=\frac{3\sigma^{J=1}_{\chi_c}}{4\pi}\biggl\{\rho^1_{0,0}
\left[r^1_0\cos^2\theta+\frac{r^1_1}{2}\sin^2\theta\right]+
\rho^1_{1,1}\left[r^1_0\sin^2\theta+
\frac{r^1_1}{2}(1+\cos^2\theta)\right]\\ \nonumber
&\qquad\qquad-\sqrt{2}\sin(2\theta)\left(r^1_0-\frac{r^1_1}{2}\right)
[\mathrm{Re}(\rho^1_{1,0})\cos\phi-\mathrm{Im}(\rho^1_{1,0})\sin\phi]\\
&\qquad\qquad-\sin^2\theta\left(r^1_0-\frac{r_1^1}{2}\right)
[\mathrm{Re}(\rho^1_{1,-1})\cos(2\phi)-\mathrm{Im}(\rho^1_{1,-1})\sin(2\phi)]\biggr\},
\label{jpsi-1}
\end{align}
where $\rho_{\lambda\lambda'}$ are the diffractive production
density matrix, $\sigma^{J=1}_{\chi_c}$ total (summed over all meson
polarisations $\lambda$) production cross section, and $r^1_0\simeq
r^1_1\simeq 0.5\,.$ Corresponding expression for tensor
$\chi_c(2^+)$ meson is much more complicated, and we do not show it
here.

Function $W(\theta,\phi)$ is a periodic one in both angles $\theta$
and $\phi$. From Eq.~(\ref{jpsi-1}) it follows that dependence on the
polar angle $\theta$ is determined mostly by diagonal terms of the
production density matrix $\rho^J_{\lambda\lambda}$, whereas
$\phi$-dependence is given by real and imaginary parts of
non-diagonal terms. Periods of oscillations in polar angle $\theta$
for $\chi_c(1^+)$ and $\chi_c(2^+)$ mesons are shifted by $\pi/2$
with respect to each other, as demonstrated in Fig.~\ref{fig:chic} (right)
for distribution $\langle W(\theta,\phi)\rangle_{\phi}$ averaged
over $\phi$.

\section{Conclusions}

We have calculated differential cross sections for central exclusive
$\chi_c(1^+,2^+)$ meson production for different spin polarisations.
The integrated cross section for the maximal helicity state is
approximately an order of magnitude greater than that for the
non-maximal ones. We have calculated, in addition, angular
distributions of $J/\psi$ meson $W^J(\theta,\phi)$ from radiative
decays $\chi_c(1^+,2^+) \to J/\psi + \gamma$. Function
$W^J(\theta,\phi)$ contains information about all independent
elements of CEP density matrix.


\begin{thebibliography}{00}

\bibitem{exp}
  T.~Aaltonen {\it et al.}  [CDF Collaboration],
  Phys.\ Rev.\ Lett.\  {\bf 102}, 242001 (2009).

\bibitem{theor}
  L.~A.~Harland-Lang, V.~A.~Khoze, M.~G.~Ryskin and W.~J.~Stirling,
  Eur.\ Phys.\ J.\  C {\bf 65}, 433 (2010);
  arXiv:1005.0695 [hep-ph].

\bibitem{KMR}
V.A. Khoze, A.D. Martin and M.G. Ryskin, Phys. Lett. B {\bf 401},
330 (1997);
Eur. Phys. J. C {\bf 23},
311 (2002).

\bibitem{chic}
  R.~S.~Pasechnik, A.~Szczurek and O.~V.~Teryaev,
  Phys.\ Lett.\  B {\bf 680}, 62 (2009);
  Phys.\ Rev.\  D {\bf 81}, 034024 (2010).

\bibitem{MR}
  A.~D.~Martin and M.~G.~Ryskin,
  Phys.\ Rev.\  D {\bf 64}, 094017 (2001).

\bibitem{Kniehl03}
  B.~A.~Kniehl, G.~Kramer and C.~P.~Palisoc,
  Phys.\ Rev.\  D {\bf 68}, 114002 (2003).

\end{thebibliography}
\end{document}